\documentclass{aa}
\usepackage{graphicx}
\begin{document}
   \title{Near-Infrared Photometry in J H and Kn bands for Polar Ring Galaxies:}

   \subtitle{II. Global Properties}

   \author{Enrichetta Iodice, 
          \inst{1,2}
          \and
           Magda Arnaboldi
          \inst{1}
          \and 
           Linda S. Sparke
          \inst{3} 
          \and 
           Kenneth C. Freeman
          \inst{4}     
          }

   \offprints{E. Iodice, \email{iodice@na.astro.it}}

   \institute{INAF-Osservatorio Astronomico di Capodimonte (OAC), 
              via Moiariello 16, I-80131 Napoli
         \and   
         International School for Advanced Studied (ISAS), 
              via Beirut 2-4, I-34014 Trieste
         \and   
          University of Wisconsin, Department of Astronomy,
             475 N. Charter St., Madison, WI
         \and  
          RSAA, Mt. Stromlo Observatory, Canberra, Cotter Road,\\ 
          Weston ACT 2611, Australia\\
             }

   \date{Received ; accepted }

   \abstract{ We discuss the properties of the host galaxy and ring light
distributions in the optical and near infrared bands for a sample of Polar
Ring Galaxies (PRGs), presented in Paper I (Iodice et al. \cite{paperI}).
The goal of this work is to test different formation scenarios for PRGs,
proposed by different authors in the last decades, by comparing their
predictions with these new data.
The strategy is twofold: {\it i)} the integrated colors of the main
components in these systems are compared with those of standard
morphological galaxy types, to investigate whether differences in colors
are caused by dust absorption or difference in stellar populations.
We then derived an estimate of the stellar population ages in 
PRGs, which can be used to set constrains on the dynamical modeling and
the time  evolution of these systems;
{\it ii)} we analyse the structural parameters of the host galaxy 
in order to understand whether this component is a standard
early-type system as its morphology suggests, and the light
distribution in the polar ring to measure its radial extension.\\
These observational results indicate that the global properties of PRGs
are better explained by dissipative merging of disks with un-equal masses
as proposed by Bekki (1998), rather than the accretion-or stripping-of 
gas by a pre-existing  early-type galaxy.

   }
   \titlerunning{NIR Photometry for PRGs}
   \authorrunning{Iodice et al.}                                                
   \maketitle
%

\section{Introduction}\label{intro}
Gravitational interactions and mergers affect the morphologies and
dynamics of galaxies. Both the Hubble Space Telescope (HST) and
the ESO Very Large Telescope (VLT) make it possible to observe the early
universe and show that these processes dominate at higher redshift
(Driver et al. \cite{driver95}).
Polar Ring Galaxies (PRGs) are considered one of the best example of
remnants from galaxy interaction. These peculiar objects are composed by a
central spheroidal component, the {\it host galaxy}, surrounded by an
outer {\it ring}, made up by gas, stars and dust, in a
nearly perpendicular plane to the equatorial one of the central galaxy
(Whitmore et al. \cite{PRC}).
The origin of PRGs is still an open debate: they may be the result of
an accretion phenomena (Toomre \cite{Toomre77}; Shane \cite{Shane80}; 
Schweizer et al. \cite{Schweizer83}; Sackett \cite{Sackett91}; Sparke
\cite{Sparke91}; Hibbard \& Mihos \cite{Hibbard95}; Reshetnikov and
Sotnikova \cite{Resh_Sot97})
or of major dissipative merger (Bekki \cite{Bekki97}; \cite{Bekki98}).
Recent observations of several interacting galaxy pairs, of similar
luminosities, display ring-like structures (e.g. NGC~7464/65,
Li \& Seaquist \cite{Li94}; NGC~3808A,B and NGC~6285/86, Reshetnikov et al. 
\cite{Resh96}).

Reshetnikov \& Sotnikova (\cite{Resh_Sot97}) studied 
the accretion scenario for the formation of PRGs using 
a smoothed-particle hydrodynamic simulations (SPH) in high speed
encounters. They analyzed the different ring morphologies
which were generated by the encounter of a gas-rich spiral with either an
elliptical or an S0 galaxy.
They followed the full history of the gas stripping: from the 
spiral galaxy outskirts to its capture by the early-type galaxy,
on a parabolic encounter. The total amount of accreted gas 
by the early-type object is about $10\%$ of the gas in the spiral
galaxy, i.e. up to $10^{9}$ $M_{\odot}$.
The size of the polar ring is found to be related to the central mass 
(luminous + dark) concentration of the host galaxy.  If the mass is highly
concentrated, i.e. the elliptical galaxy case, the ring forms at smaller
radii; if the host galaxy has an extended massive halo, i.e. the S0 case,
the ring average radius ($\bar{R}$) can be as large as 30~kpc.
This scenario can account for the formation of (quasi-)stable polar rings, 
whose radial extent is of the order of 10\% of the ring extention.

A quite different approach to the formation of polar ring galaxies was recently
proposed by Bekki (\cite{Bekki98}). In this scenario, the polar ring
results from a ``polar'' merger of two disk galaxies with unequal mass.
The ``intruder'', on a polar orbit with respect to the ``victim'' disk,
passes through it near its center: it is slowed down, and pulled back
toward the victim, by strong dissipation which is caused by the interaction
with the victim gaseous disk.                            
Dissipation removes random kinetic energy from the
gaseous component of the victim's disk, so that some gas
can settle again into a disky configuration.
The morphology of the merger remnants depends on the merging
initial orbital parameters and the initial mass ratio of
the two galaxies.
                                                               
Bekki's scenario successfully reproduces the range of observed morphologies
for polar ring galaxies, such as the existence of both
wide and narrow rings, helical rings and double rings  (Whitmore
\cite{whitmore91}).
Furthermore, this scenario would also explain the presence of wide and massive 
polar disk, as observed in NGC~4650A (Arnaboldi et al. \cite{magda97}, 
Iodice et al. \cite{4650aI}; Gallagher et al. \cite{4650aG}).
The two scenarios, accretion vs. dissipative mergers of disks, both
predict the general features of PRGs: a structure-less appearance of the
host galaxy, 
and the younger dustier appearance of the polar structure. 
But they differ on their predictions about structural parameters, age,
baryonic mass and polar structure extension. Therefore, the two
scenarios should be tested against the observed
properties of both  wide and narrow PRGs, in particular their observed 
structural parameters, colors and total light. 
Given that PRGs contain a lot of dust, we
must study their light distribution in the near-infrared (NIR), and
determine the distribution of their evolved stellar population.
To this aim, new NIR data for a sample of PRGs were collected and analysed
in Iodice et al. \cite{paperI}, hereafter Paper I.
In this work we compare the integrated colors derived for the host galaxy 
and ring of each PRG in our sample (see Sec.5 in Paper I) with those of  
standard morphological galaxy types, in Sec.\ref{nircol},
and compute an estimate of the stellar population ages, in Sec.\ref{PR_age}. 
In Sec.~\ref{model} and Sec.~\ref{param_pr}, we perform a detailed
analysis of the light distribution properties in the 
host galaxy and ring.  The new observational evidences obtained for this
sample of PRGs are summarized in Sec.~\ref{sum}, and conclusions are 
derived in Sec.~\ref{conclu}.

\section{NIR colors}\label{nircol}
The near-infrared J, H and Kn images for a sample of PRGs were obtained at
the 2.3 m telescope of the Mt. Stromlo and Siding Spring Observatory, with
the CASPIR camera; all PRGs in our sample are listed in Table~\ref{colcode}.
A detailed description of the observations and data reduction were given
in Paper I (see Sec.~2). For each object, we computed the
integrated magnitudes in J, H and Kn bands, in five different areas 
(shown in Fig.~8, Paper~I), which cover the nucleus of the system, 
the host galaxy stellar component (outside the nucleus), and the polar 
structure (see Paper I, Sec.5 for more details). 

\begin{figure*} 
\centering
\includegraphics[width=15cm]{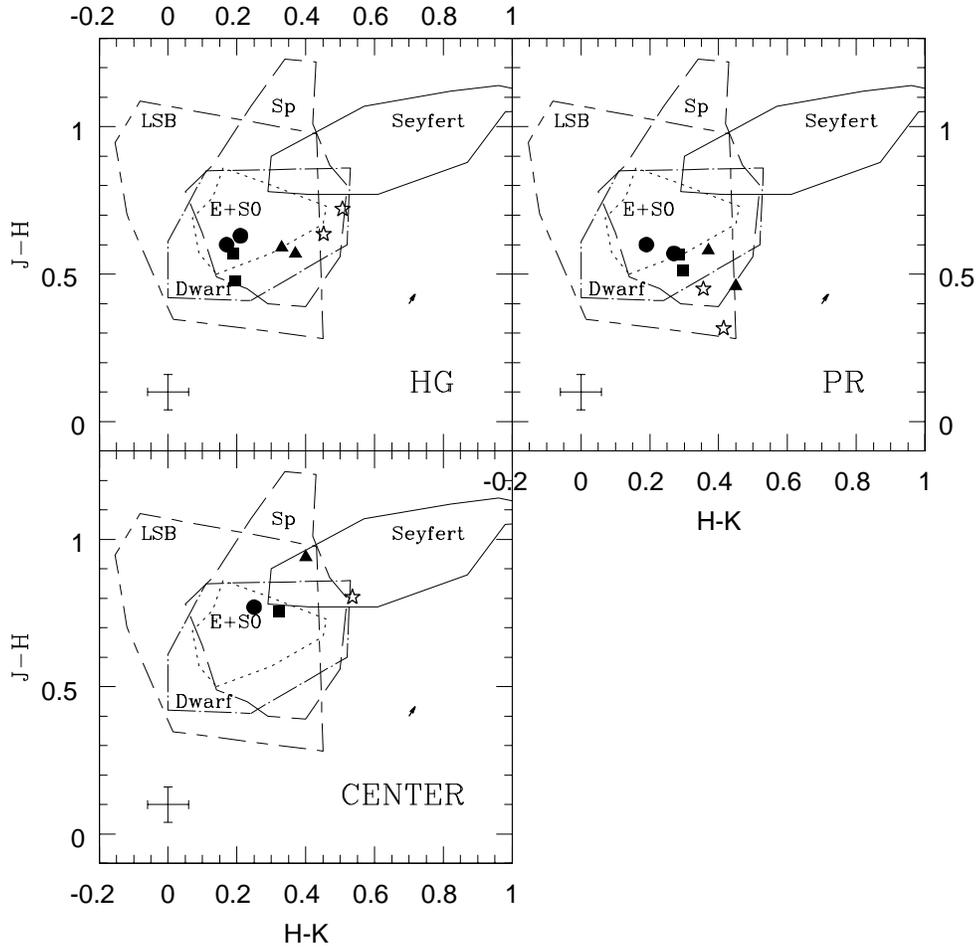}  
\caption{J-H vs. H-K color diagram for the two regions of the host galaxy  
(top left panel), two regions for the polar ring (top right panel) and for  
the central area  (bottom left panel) in all PRGs of our sample. Code for
symbols is: stars for ESO~415-G26, filled squares for ARP~230, filled
circles for AM~2020-504 and  filled triangles for ESO~603-G21.
The dotted contour limits the region where the integrated colors of Es and S0s
are found; the long-dashed contour limits the region where the integrated colors of 
spirals are found;
the dashed-dotted contour identifies the integrated colors  of the dwarf 
elliptical galaxies; the long dashed - short dashed contour identifies the 
integrated colors of LSB galaxies, and the continuous line limits the 
integrated colors of the nuclear regions of Seyfert 1 and 2 galaxies.     
The arrow, in the lower right corner, indicates the reddening vector direction
computed for galactic dust in the screen model approximation, 
see discussion in Sec.\ref{nircol}. The average errors are shown in the lower
left corner.}                
\label{jhkplot}
\end{figure*}

We then derived the J-H vs. H-K integrated colors 
for the host galaxy and ring, see Paper I, Sec.~5.
We now compare them with those of (1) standard early-type galaxies in the
Fornax and Virgo clusters (Persson et al., \cite{Persson79}), (2) spirals
(Giovanardi \& Hunt, \cite{Giov96};
Frogel, \cite{Frog85}; de Jong \& van der Kruit, \cite{deJong94}), 
(3) dwarf ellipticals (Thuan, \cite{Thuan85}), (4) low surface brightness  
galaxies, LSB, (Bergvall et al. \cite{Berg99}), and with the inner regions
of Seyfert 1 and 2 (Glass and Moorwood, \cite{Glass85}), see Fig.\ref{jhkplot}.
In all PRGs of our sample, the host galaxy has on average bluer colors
than the typical values for early-type galaxies. They are more 
similar to the colors of spiral and dwarf galaxies, with the exception of 
AM~2020-504. This component is also characterized by a strong color
gradient toward bluer colors in J-H, from the central regions 
going outwards. The outer regions have on average similar H-K colors. \\
In the screen model approximation, we can compute
the reddening vector by assuming $A_{V}=0.3$, 
as in the Milky Way galaxy (Gordon, Calzetti \& Witt \cite{GCW97}); 
the result is shown in Fig.\ref{jhkplot}.
It suggests that the dust reddening is small in these bands, 
and it can account for the observed color gradient between nucleus and
stellar component in the host galaxy, with the exception of ESO~603-G21,
and within the uncertainties of the color estimate. 
The central regions of ESO~603-G21 have colors which are typical of a
Seyfert galaxy, as it was found by Arnaboldi et al. in
1995. Similar behavior is observed for the nuclear region of ESO~415-G26.\\
In almost all PRGs, the polar structure has on average bluer colors than 
the host galaxy, and quite similar to those of the late-type
galaxies.

We then have derived the B-H vs. J-K integrated colors for the 
host galaxy and ring, which are shown in Fig.\ref{age1}. The host galaxy
in all PRGs of the sample, but AM~2020-504, has
overall bluer B-H colors than the average values observed in early-type
galaxies (Bothun \& Gregg \cite{Bothun90}); 
on average, they are very similar to those of spirals (by Bothun et al., 
\cite{Bothun84}). 
We can then expect a younger age for the central component of PRGs
than those predicted for early-type systems.
In nearly all objects, one side of the host galaxy is bluer than the other
side: this is most likely caused by the presence of the polar ring, which
perturbs the regions where it passes in front of the galaxy, along the
line-of-sight.

As was already found in the J-H vs H-K diagram, there is
a strong color gradient between the central region of the host galaxy and its
outer parts, for all PRGs of the sample.  
The very red B-H colors of nuclear regions may be due to the dust absorption:
as extreme examples, in the central regions of ARP~230
most of light is completely obscured by dust in the optical band.
The reddening vector, shown on Fig.\ref{age1}, is also computed for these 
color indices: the absorption due to the dust may account for
the color gradient in the host galaxy of almost all PRGs.

\begin{table}
\caption[]{PRG sample.}
\label{colcode}
\begin{tabular}{cccc}
\hline\hline
Object & $\alpha$ (J2000) & $\delta$ (J2000) & Filter\\
\hline
A0136-0801& 01h38m55.2s & -07d45m56s & H \\
ESO 415-G26 & 02h28m20.1s & -31d52m51s & JHKn \\
ARP 230 & 00h46m24.2s & -13d26m32s & JHKn \\
AM 2020-504& 20h23m54.8s & -50d39m05s & JHKn \\
ESO 603-G21& 22h51m22.0s & -20d14m51s & JHKn \\
\hline
\end{tabular}
\end{table}

\section{Using colors to date the stellar populations in PRGs}\label{PR_age}
The next step is to compare the B-H vs. J-K integrated colors
with the stellar population models.
The goal is to derive the ages of the dominant stellar
populations in the central spheroid and polar structure.
The B-H and J-K colors are
used to break the age-metallicity degeneracy, as suggested by Bothun et
al. (\cite{Bothun84}). The J-K color gives a good estimate
of the metallicity and it is quite insensitive to the presence of a young
stellar population. 
This effect is also supported by the observed monotonic increase
of the mean J-K color in globular clusters with increasing metallicity
(Aaronson et al. \cite{Aar78}, Frogel et al. \cite{Frog83}). In addition, the
population synthesis models by Bothun (\cite{Bothun82}) show that
J-K is decreased only by $0.05$ mag as a result of a star burst,
while the B luminosity is increased by $1$ mag by the same stellar burst.
The B-H color is sensitive to the combined effect of the Star Formation Rate 
(SFR), metallicity and age (Bothun et al. \cite{Bothun84}).

The stellar population synthesis model developed by Bruzual \& Charlot 
(\cite{BC93}), GISSEL ({\it Galaxies Isochrone Synthesis Spectral Evolution 
Library}) was used to reproduce the integrated colors of different regions
(see Sec.\ref{nircol}) in each PRGs of the sample. 
We first select a set of models which were able to reproduce
on average the integrated colors of galaxies with different morphological
types in the local Universe. These models were then optimized to reproduce the
observed colors for the main components of each PRG.
The GISSEL key input parameters are the Initial Mass Function (IMF),
the Star Formation Rate (SFR), and the metallicity. 
In what follows, we have assumed that stars form according to the
Salpeter (\cite{Salp55}) IMF, in the range from $0.1$ to $125 M_\odot$.

\begin{figure*} 
\includegraphics[width=15cm]{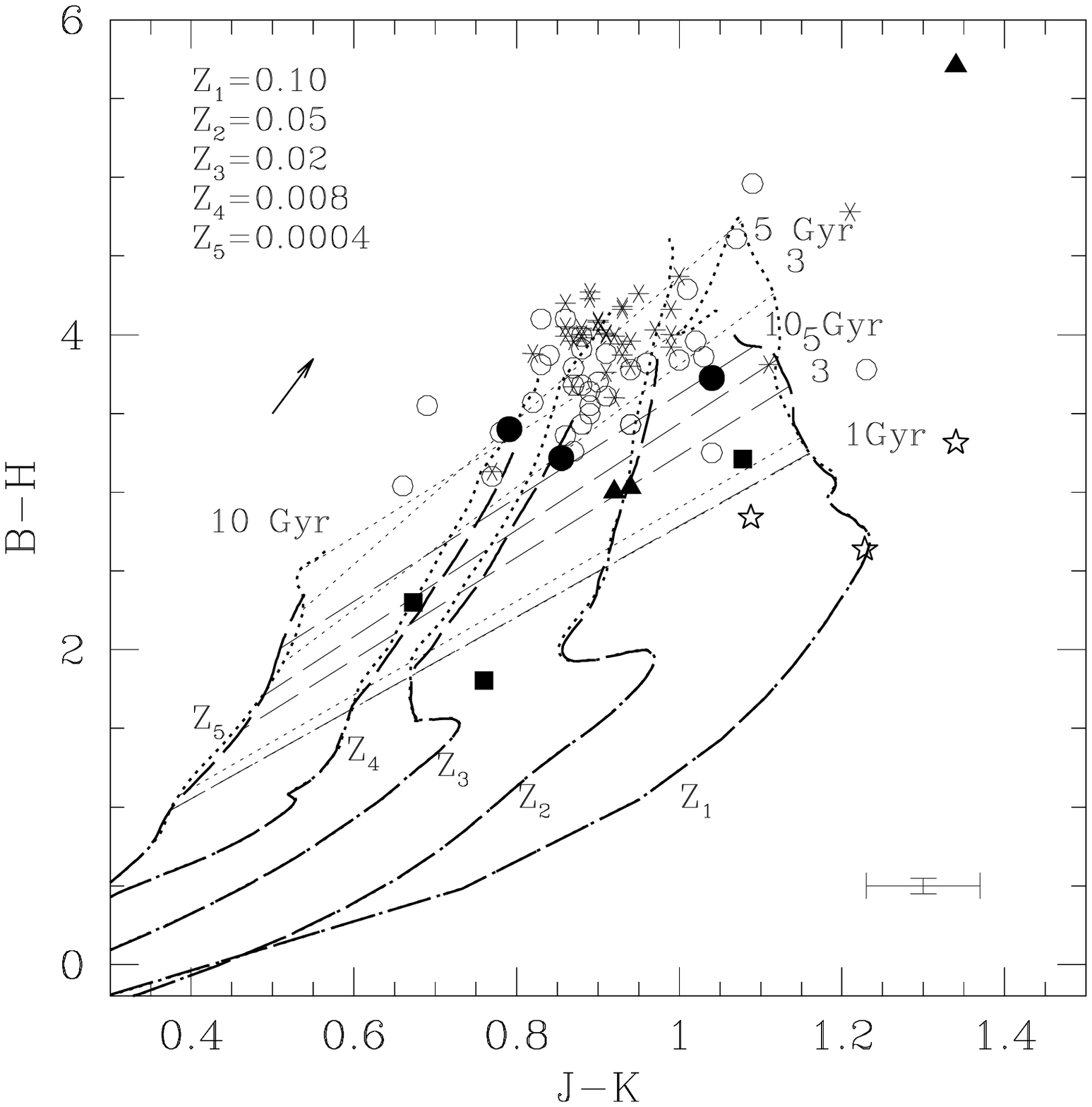}
\caption{B-H vs. J-K color diagram of the evolutionary tracks for the
stellar synthesis models optimized for the host galaxy, for each PRG of the 
sample. Code for symbols is the same adopted in Fig.\ref{jhkplot}. 
The arrow (on the left side) indicates the reddening vector direction 
for galactic dust and the screen model approximation, 
quoted in Sec.5 of Paper~I. The average errors are indicates in the bottom 
right side.
The heavier dotted lines correspond to models with a characteristic
timescale $\tau=1 Gyr$ and heavier dashed lines for models with $\tau=7 Gyr$.
Models are computed for different metallicities as shown on this figure.
Light dotted and light dashed lines indicate loci of constant age for the
different models; different ages are reported on the plot.
Open circles and asterisks correspond to bulges and disks respectively 
from a sample of S$0$ galaxies (Bothun \& Gregg \cite{Bothun90}).}
\label{age1}
\end{figure*}

To derive the age estimate for the polar structure and host galaxy, 
different evolutionary models are adopted, given by different SFRs.
A star formation history with an exponentially decreasing rate,
given by $SFR(t)= \frac{1}{\tau} \exp{(- t/ \tau)}$,
was adopted for the central host galaxy. 
The $\tau$ parameter quantifies the ``time scale'' when the star formation
was most efficient. The adopted expression for the SFR is used
to reproduce the photometric properties of the elliptical galaxies,
and can be derived from the assumption that the rate with which stars form is
proportional to the available gas quantity (Kennicutt, \cite{Kenn83}).
In order to obtain the largest range for the age estimate, the following
two values were adopted for the time scale parameter: 
$\tau=1$ Gyr and $\tau =7$ Gyr. Each model for the host galaxy 
were derived for the following metallicity values: $Z=0.0004$, $Z=0.008$, 
$Z=0.02$, $Z=0.05$, $Z=0.1$, which are constant with age.
The corresponding evolutionary tracks were derived for each
metallicity and are plotted in Fig.~\ref{age1} with 
the lines of constant age.
 
The photometric properties of a sample of early-type galaxies by Bothun et al. 
(\cite{Bothun84}) are well reproduced with a $\tau =1$~Gyr model, which
predicts an age of about $10$ Gyr for the dominant stellar population in
these systems.
On the other hand, the same model predicts younger ages, between $1$ to
$3$ Gyr, for the host galaxy (nucleus and outer stellar component) in all
PRGs of this sample. The exception is AM~2020-504, which seems to be
as old as standard early-type galaxies, 
see Fig.~\ref{age1}.
For the sample of early-type galaxies and the polar ring galaxy AM~2020-504, 
the model with $\tau =7$ Gyr implies an age older than $10$ Gyr, 
whereas the other PRGs of the sample have younger age, between $1$ to
$3$ Gyr, see Fig.~\ref{age1}.

\begin{figure*} 
\includegraphics[width=15cm]{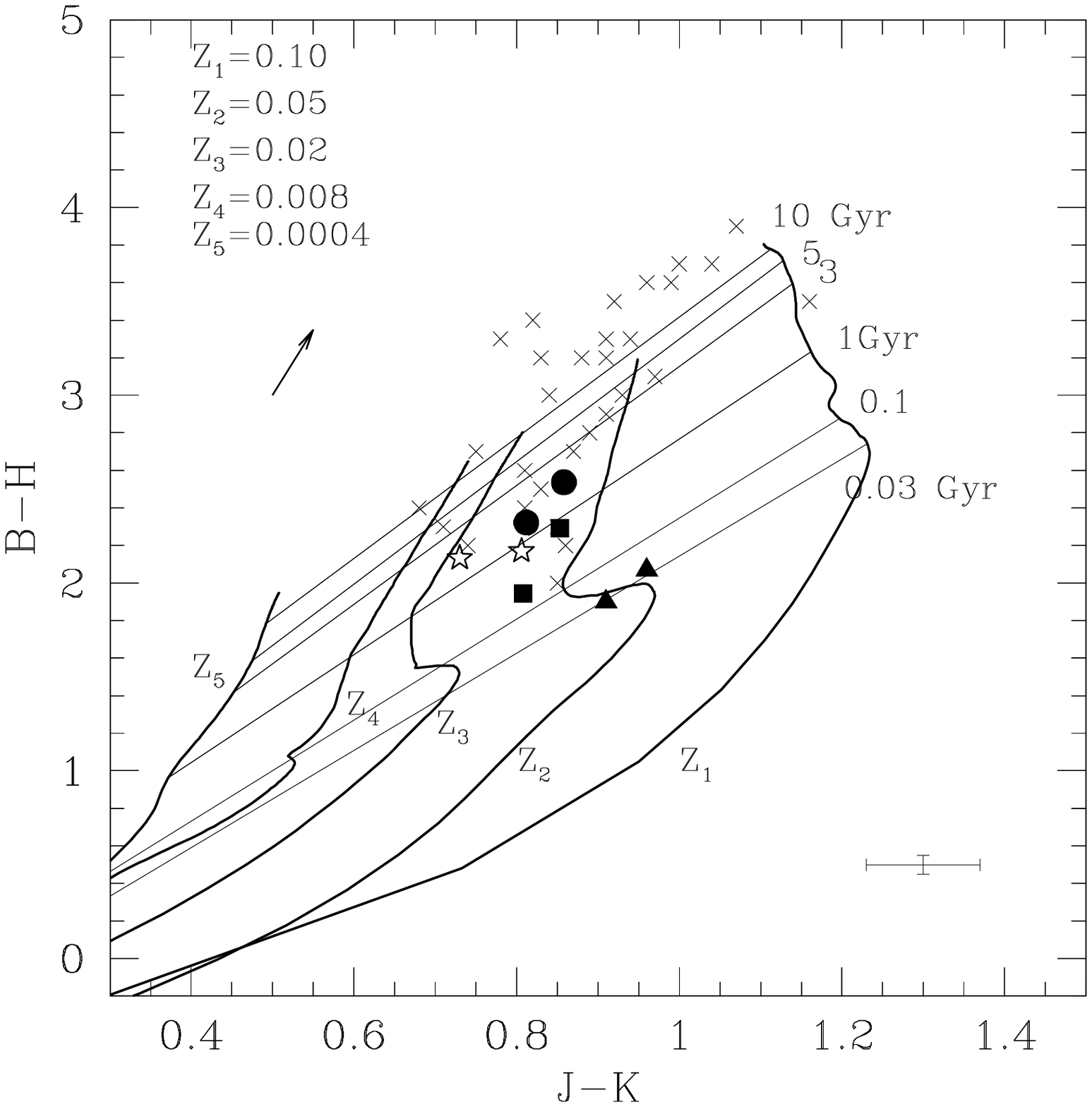}
\caption{B-H vs. J-K color diagram of the evolutionary tracks for the
stellar synthesis models optimized for the polar structure, 
for each PRG of the sample. 
Code for symbols is the same adopted in Fig.\ref{jhkplot}.
The arrow (on the left side) indicates the reddening vector direction 
for galactic dust and the screen model approximation, quoted in Sec.5 of 
Paper I. 
The average errors are indicates in the bottom right side. Heavier lines
indicate models  with constant SFR, computed for different metallicities
(reported on the plot).
Light lines are loci of constant age; different ages are quoted on the plot. 
Crosses are for a sample of spiral galaxies (Bothun et al., \cite{Bothun84}).}
\label{age2}
\end{figure*}

On average, the polar structure has bluer colors than the
host galaxy, which suggests even a younger age for this component.
Observations of HII regions and blue star clusters associated with the
polar component suggest that star formation is active in polar rings, 
so a constant star formation rate, 
$ SFR(t)= K$ with $K \sim 10^{-10}$ $M_{\odot}/yr$, 
(with metallicities $Z=0.0004$, $Z=0.008$, $Z=0.02$, $Z=0.05$, $Z=0.1$) 
was adopted. For each metallicity,
the corresponding evolutionary tracks were obtained and the lines of 
constant age derived, see Fig.~\ref{age2}.
These models successfully reproduce the mean colors for a sample of spiral
galaxies (Bothun et al., \cite{Bothun84}) and 
imply an average age of about $5$ Gyr for these objects.
The integrated colors of the polar structure in all PRGs of this sample are 
very similar to the bluer/younger spiral galaxies; they are also clustered 
in the same metallicity range, between $Z=0.02$ and
$Z=0.05$, and similar age of about $1$ Gyr. 

We wish to stress that the colors derived for both components were not
corrected for the intrinsic reddening due to the PR system , i.e. the
absorption caused by the dust in the polar structure; therefore the true
colors of the stellar population associated with the central host galaxy 
might be even bluer. 
Furthermore, the age estimates for the host galaxy and
the polar structure are uncertain, because we do not have independent
information on the star formation law and metallicity
of the stellar population in the two components.
The intrinsic uncertainties of the synthesis population models must
also be considered, particularly for the age of the host galaxy.
By comparing three recent synthesis codes, Charlot, Worthey and
Bressan (\cite{CWB96}) found that the colors predicted for old populations with
an age $>1$ Gyr, with the same input age and metallicity, are affected by
discrepancies, which are primarily due to the different prescriptions
adopted for the stellar evolution theory.
Thus, the age estimates given here, for the central host galaxy and 
polar structure, should be considered only indicative.

\section{Study of the host galaxy light distribution}\label{model}
One of the open issues in the study of PRGs is the nature
of the host galaxy. A qualitative morphological inspection suggests that
this component is similar to an early-type galaxy, most often an S0 
(see Paper~I, Sec.~3). 
However, the NIR integrated colors, derived for the host galaxy
in this work (Sec.\ref{nircol}) and in previous ones (Arnaboldi et al. 
\cite{magda95}, Iodice et al. \cite{4650aI}), are bluer on average than those 
for standard early-type galaxies.
The study of the surface brightness light distribution in the host galaxy may 
represent an independent way to check whether this component is really an 
S0 galaxy or not, by investigating how the structural parameters compare with 
those of standard early-type galaxies.  

\subsection{Bulge effective parameters: $\mu_{e}$ - $\log r_{e}$ 
plane}\label{param_hg1}
In Fig.~\ref{bulge} (left panel) we plot the structural parameters of PRG  
host galaxies in the $\mu_{e}$ and $r_{e}$ plane.
We select a sample of S0 and spiral galaxies for the comparison. 
The sample of early-type galaxies were studied by Bothun and Gregg (in 1990).
The structural parameters for these
objects where obtained by fitting the light distribution with the 
super-position of a de Vaucouleurs law ($n=4$) and an exponential disk, 
for the bulge and disk component respectively, in the B band. 
The surface brightness parameters in the K band are derived by considering 
the average B-K color of these objects (Bothun and Gregg, \cite{Bothun90}).
The sample of spiral galaxies include objects studied by M\"{o}llenhoff 
\& Heidt (\cite{mollenhoff2001}) and by Khosroshahi et al. (\cite{khos2000b}). 
The host galaxy of PRGs are characterized by a more ``compact'' bulge with  
respect to normal early-type galaxies, except for AM~2020-504 which 
has larger values of $\mu_{e}$ and $r_{e}$. 
Reshetnikov et al. (1994) found a similar behavior for the bulges
of polar ring galaxies in the B band. Because PRGs have smaller bulges
also in the NIR, where the perturbations due to the dust absorption are
minimal, we know now that the PRG bulges are indeed
smaller\footnote{ Reshetnikov et al. (\cite{Resh94}) were not able to
exclude the hypothesis that the bulge luminosities and sizes were
underestimated because of dust absorption by the polar ring.}.
The 2D bulge-disk decomposition adopted here (see Sec.6, Paper I) 
allow us to minimize the perturbations due to the ring and to conclude 
that the bulge component of the host galaxy in PRGs resembles the smallest
early-type galaxies with the higher surface brightness in the $\mu_{e}$
- $\log r_{e}$ plane.

\subsection{Bulge scale parameters: $\log n$ - $\log r_{e}$ 
plane}\label{param_hg2}
\begin{figure*}
\includegraphics[width=8cm]{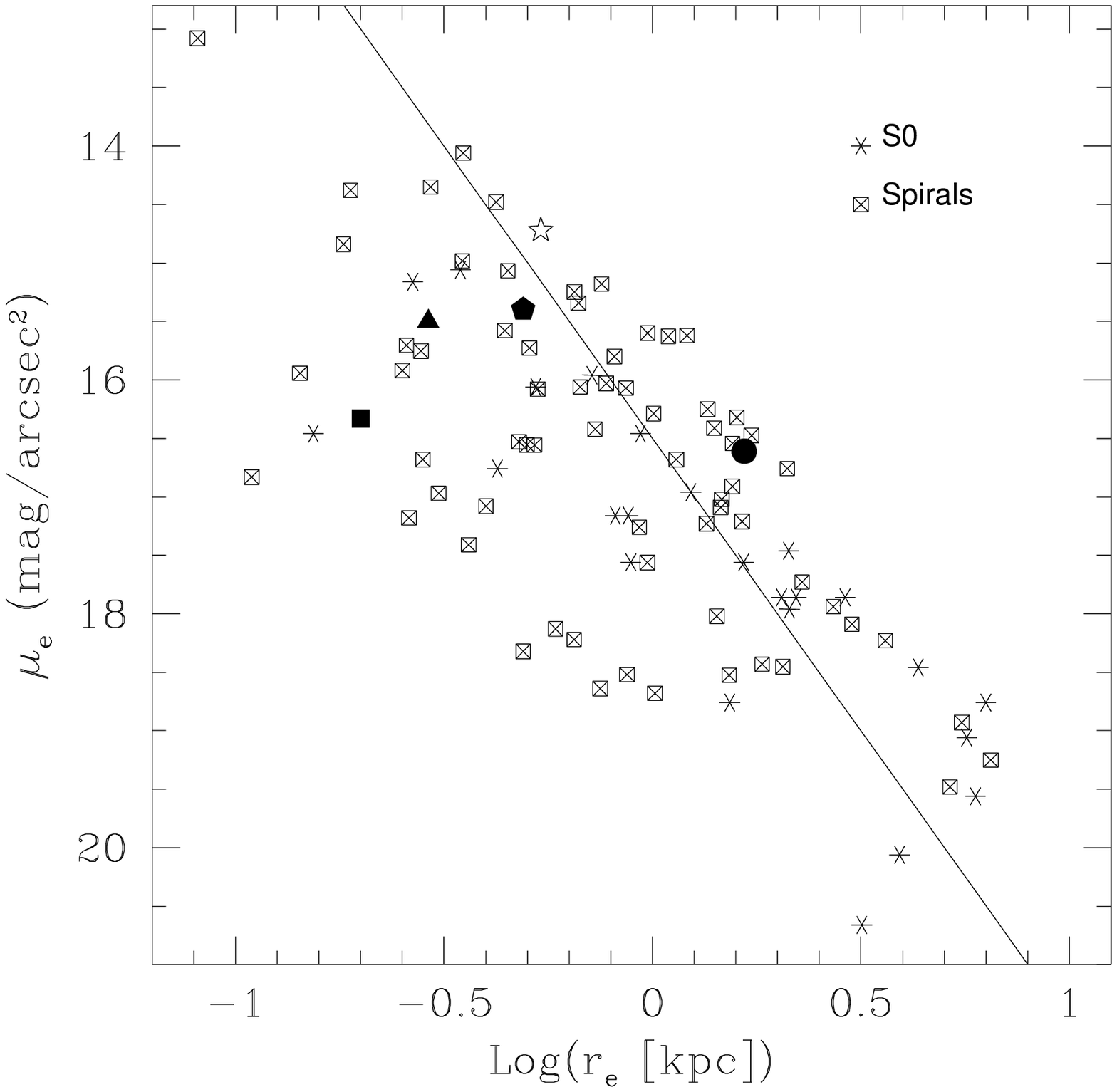}
\includegraphics[width=8cm]{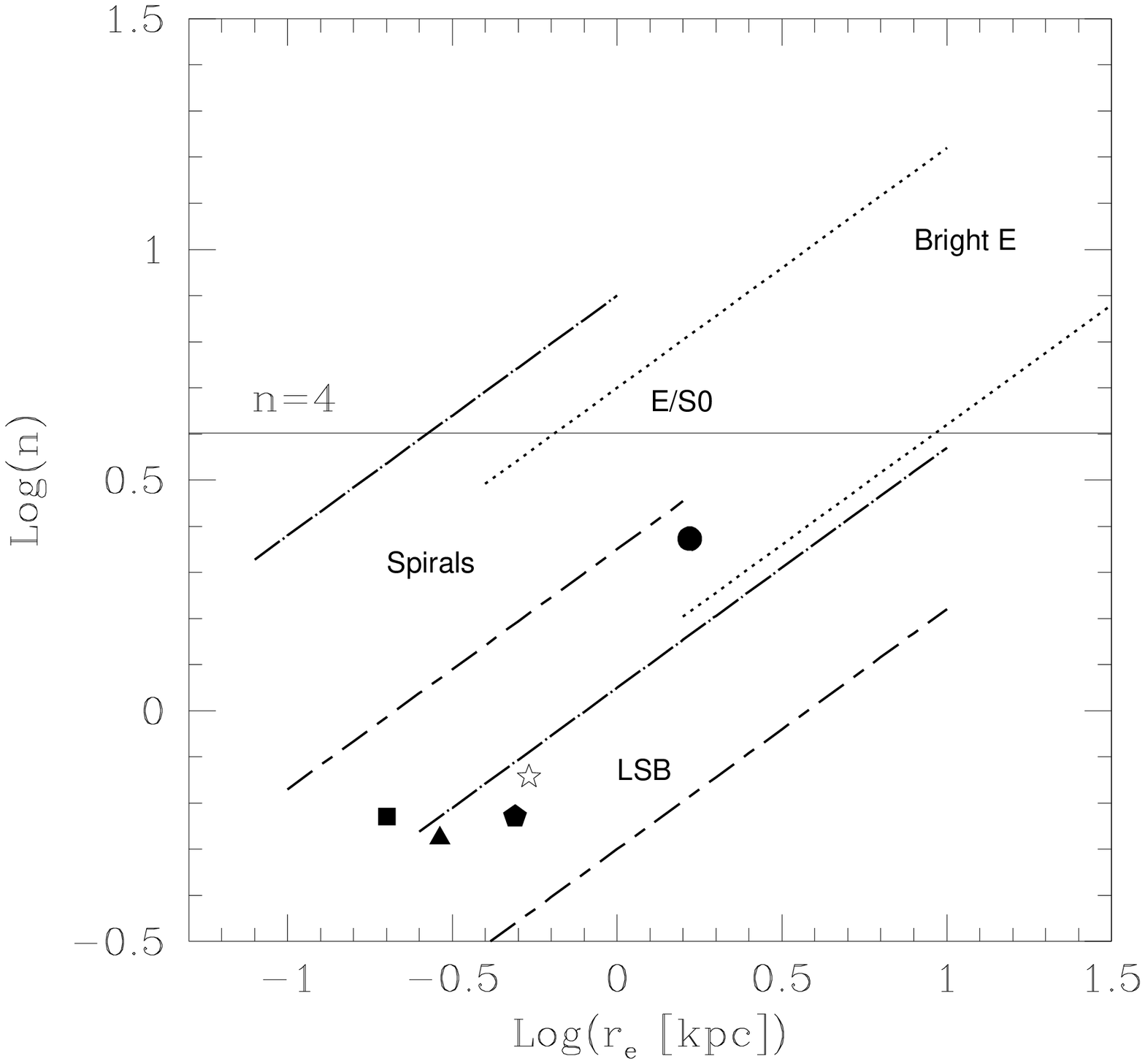}  
\caption{Left panel - Relation between the bulge effective parameters 
(see Sec.6 in Paper I) for all PRGs in our sample. 
Code for symbols is: star for ESO~415-G26, filled square for ARP~230, filled
circle for AM~2020-504, filled triangle for ESO~603-G21 and filled pentagon
for A0136-0801. They are compared with the typical values for early-type 
galaxies (Bothun and Gregg, \cite{Bothun90}) and for spiral galaxies
(M\"{o}llenhoff \& Heidt \cite{mollenhoff2001}; Khosroshahi et al. 
\cite{khos2000b}).
The solid line is a line of constant bulge luminosity derived for 
$\mu_{e}=18$ $mag/arcsec^{2}$, $r_{e}=2$ kpc and $n=4$.
Right panel - Relation between the effective radius  $r_{e}$ and the $n$
exponent of the generalized de Vaucouleurs' law (see Sec.6, Paper I), 
for the bulge component in the PRGs of our sample.
They are compared with the typical values for 1) early-type galaxies, 
Ellipticals and S0s (region limited by dotted lines), by Caon et al. 
(\cite{Caon93}); for 2) LSB galaxies (region limited by dashed lines), by 
Davies et al. (\cite{Davies88}); and for 3) spiral galaxies 
(region limited by dashed-dotted lines) by M\"{o}llenhoff \& Heidt 
(\cite{mollenhoff2001}) and Khosroshahi et al. (\cite{khos2000b}).
The plotted $n$ exponents values for the early-type galaxies are those derived
by Caon et al. (\cite{Caon93}) along the minor axis of the system,
in order to exclude the contribution  from a possible disk component.} 
\label{bulge}
\end{figure*}

The optical light distribution in the bulge of the host galaxy has a
quasi-exponential behavior, as suggested by the small values of the $n$ 
exponent (close to 1), in all PRGs of the sample, except for AM 2020-504 
(see also Table~6, Paper I). 
In Fig.~\ref{bulge} (right panel) we show the comparison between the scale 
parameters  $n$ and $ r_{e}$ for the bulge component in PRGs, with those
for a sample of early-type galaxies in the  Virgo cluster (Caon et al.,
\cite{Caon93}), for spiral galaxies (M\"{o}llenhoff \& Heidt 
\cite{mollenhoff2001}; Khosroshahi et al. \cite{khos2000b}) and
for a sample of early type 
galaxies of low surface brightness (Davies et al. \cite{Davies88}) 
in the Fornax cluster. Since Davies et al. (\cite{Davies88}) fit the B 
band light profiles by adopting the generalized de Vaucouleurs law in the
form  $I(r)=I_{0}exp[-(r/A)^{N}]$, we have computed the effective radius
$r_{e}$  and the exponent $n$ from this scalelenght radius $A$ and the
exponent $N$. 

In this plane, PRGs fall in the region where lower values for both
parameters are found, together with the LBS and spiral galaxies.
The PRGs bulge scale parameters seem to ``follow'' the relation between
the $n$ exponent and the effective radius, found by Caon et al. 
(\cite{Caon93}), for which $n$ increases steadily with  $r_{e}$. 
The polar ring galaxy AM~2020-504 is the only object of the sample which
falls  in the same regions occupied by ``ordinary family''
of early-type galaxies (Capaccioli et al. (\cite{Cap92}).

In their NIR study of spiral galaxies M\"{o}llenhoff \& Heidt 
(\cite{mollenhoff2001}) have shown that the {\it n} exponent and effective 
radius
correlates with the Hubble types, and that late-type spirals are 
characterized by the lowest values of the {\it n} exponent and have the 
smaller bulges. The PRG bulge scale parameters correspond to those of the 
late-type galaxies.
However, the shape parameter {\it n} derived for PRGs must be considered 
as a lower limit: we have found
that the light in PRGs bulges is very concentrated toward the center, so
the convolution of such light distribution  with the PSF may cause a
smoothing of the light profiles toward the center, leading to a biased
small value for the {\it n} exponent.

\subsection{Photometric plane for bulges}
Khosroshahi et al. (\cite{khos2000a}) found that a single photometric plane
exists in the space of structural parameters describing both Ellipticals and 
bulges of early-type spiral galaxies. 
They showed that $log n$, $log r_{eff}$ and the central surface brightness
$\mu _b(0)$ are related by the equation 
$log n=(0.172 \pm 0.020)log r_{eff} -(0.069 \pm 0.004)\mu_ b(0)+(1.18 \pm 0.04)$,
which is the best-fit plane for elliptical galaxies and the bulges of spiral 
galaxies. The photometric plane was used to constrain the bulge 
formation mechanisms in galaxies.
M\"{o}llenhoff \& Heidt (\cite{mollenhoff2001}) found that 
late-type disk galaxies also share this single plane, and they stressed that
this is a further hint to a common mechanism for bulge formation in all type 
of galaxies. 

In the left panel of  Fig.~\ref{disk} we plot the photometric plane for
bulges of spiral galaxies (by Khosroshahi et al. \cite{khos2000b} and 
M\"{o}llenhoff \& Heidt \cite{mollenhoff2001}) for the host galaxy in PRGs
of our sample. AM~2020-504 is the only PRG which shares the same plane of
spiral and elliptical galaxies galaxies. The other PRGs of the sample fall
away from this plane, in the lower-left corner, which is not as widely populated 
by the standard type of galaxies.

\subsection{Disk structural parameters: $\mu_{0}$ - $\log r_{h}$ 
plane}\label{param_hg3}
\begin{figure*}
\includegraphics[width=8cm]{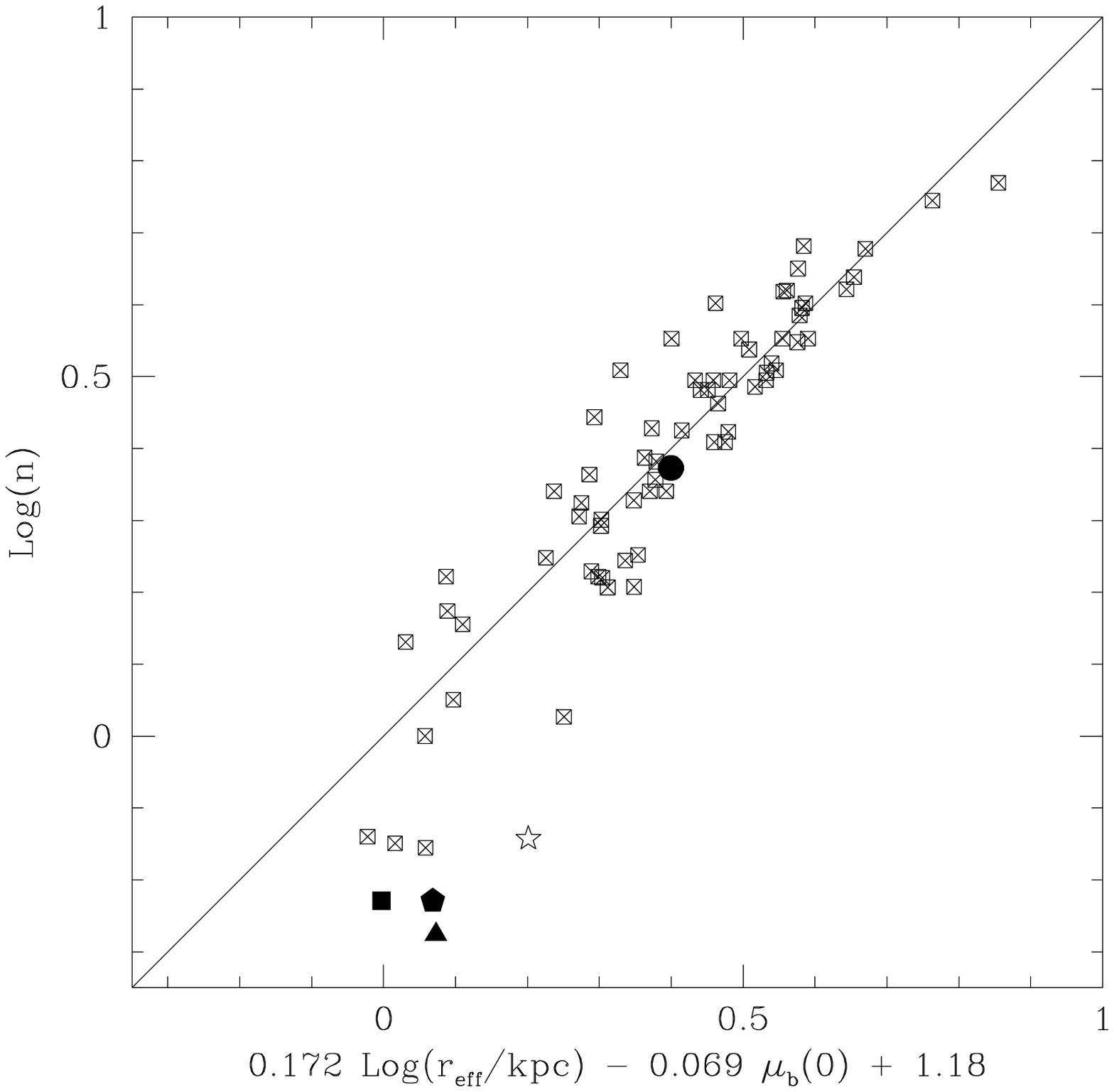}  
\includegraphics[width=8cm]{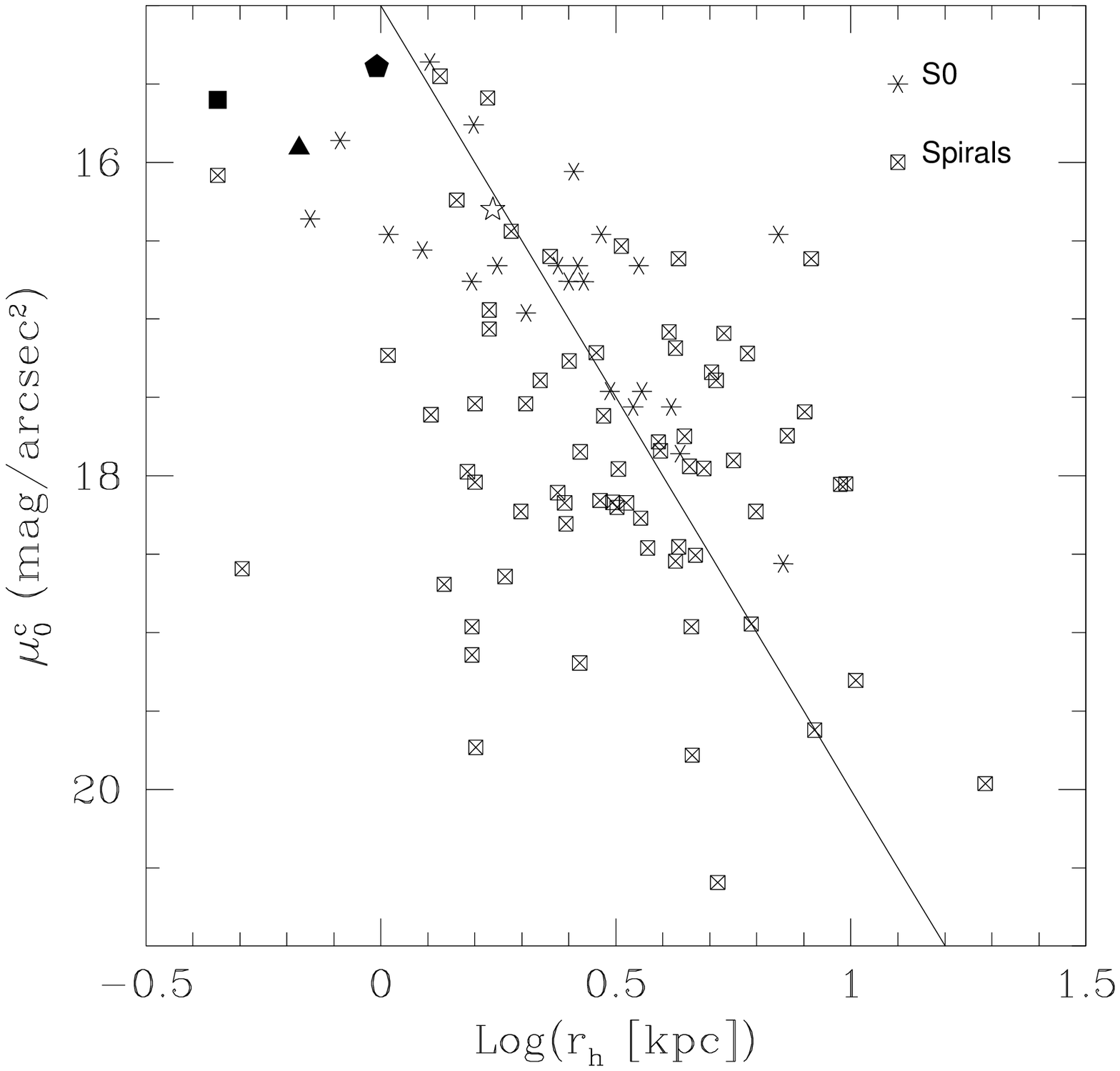}
\caption{Left panel - Photometric plane for bulge of spiral galaxies 
and for PRGs.
Right panel - Relation between the central surface brightness,  
corrected for the inclination, and the scalelength of the disk component
in the host galaxy (see Sec.6 in Paper~I).
Code for symbols is the same adopted in Fig.\ref{bulge}.
They are compared with the typical values for S0 galaxies by Bothun and 
Gregg (\cite{Bothun90}) and for spiral galaxies
(M\"{o}llenhoff \& Heidt \cite{mollenhoff2001}; Khosroshahi et al. 
\cite{khos2000b}). The solid line is a line of constant disk luminosity, 
derived for $\mu_{0}^c=18$ $mag/arcsec^2$ and $r_{h}=1.6$ kpc.}
\label{disk}
\end{figure*}

The average value of the central surface brightness (corrected
for the inclination, see Sec.6, Paper I) for this sample of PRGs is 
$<\mu_{0}^{c}>=15.9 \pm 0.3$ mag arcsec$^{-2}$. This value is 
brighter than the average value for the S0 galaxies 
(by Bothun and Gregg\footnote{The central surface brightness in the 
K band for the Bothun and Gregg (\cite{Bothun90}) sample of S0 is computed 
from their values in the B band and the B-K color of these galaxies.}
 \cite{Bothun90}), which is $<\mu_{0}^{c}>=16.6 \pm 0.9$ mag arcsec$^{-2}$, and
for spiral galaxies (by M\"{o}llenhoff \& Heidt \cite{mollenhoff2001}), 
which is $<\mu_{0}^{c}>=18 \pm 1$ mag arcsec$^{-2}$.
The average scalelength of the disks in the host galaxy is  
$<r_{h}>=1.0\pm 0.6$ kpc, whereas the average value for this sample of S0 
galaxies is $<r_{h}>=3 \pm 1$ kpc and for spiral galaxies is 
$<r_{h}>=4.5 \pm 2$ kpc.
Disks in the PRG host galaxy are brighter and smaller than the
disks of S0 galaxies: this is clearly evident in
Fig.~\ref{disk} (right panel), which shows the location of the PRGs and 
S0 disks parameters (see also Table~6, in Paper I) in the 
$\mu_{0}^{c}$ - $\log r_{h}$ plane. 
The same result was found by Reshetnikov et al. \cite{Resh94} for
a sample of PRGs in the B band.

\subsection{Bulge-Disk relations}
\begin{figure*}
\includegraphics[width=8cm]{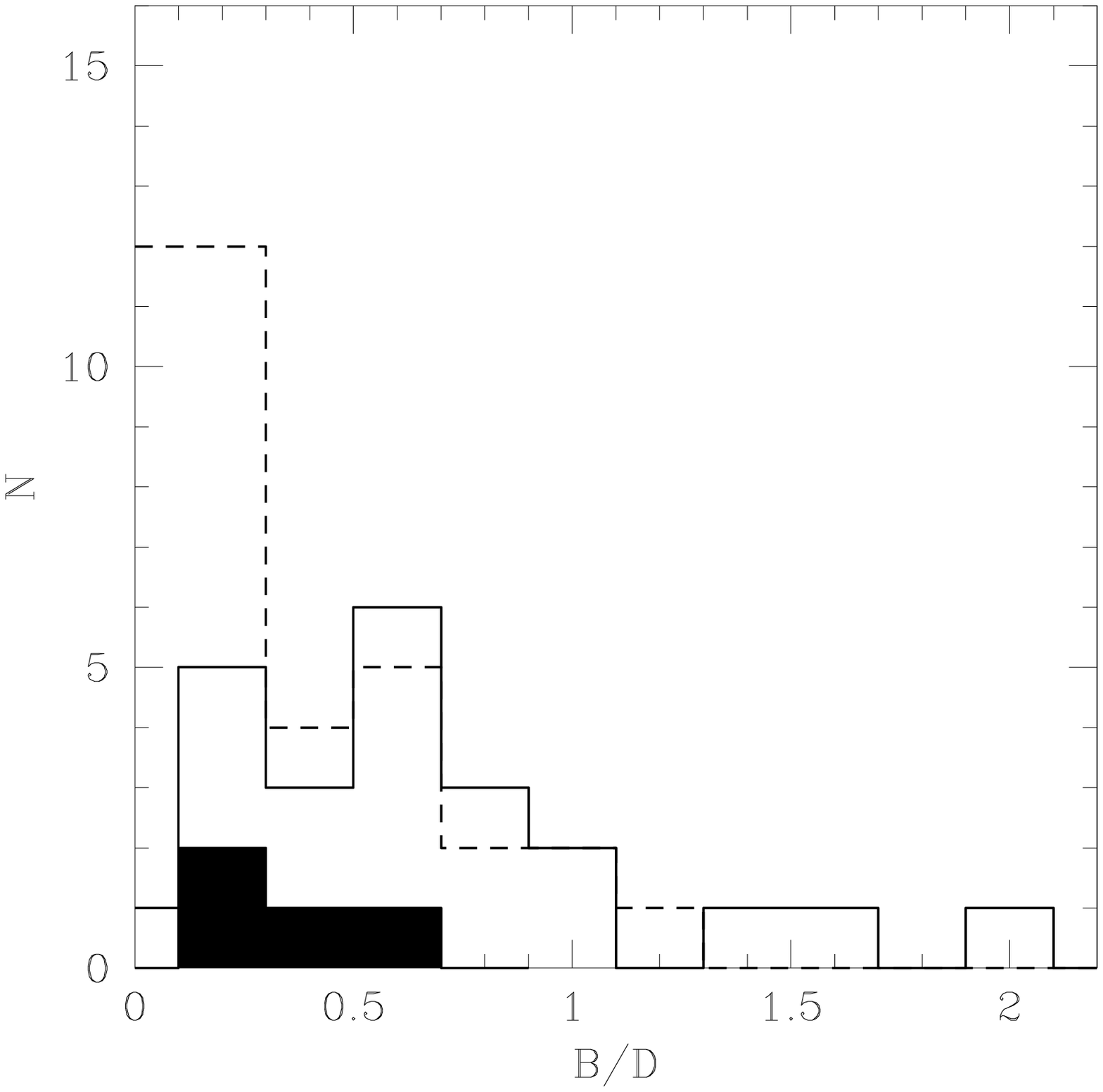}
\includegraphics[width=8cm]{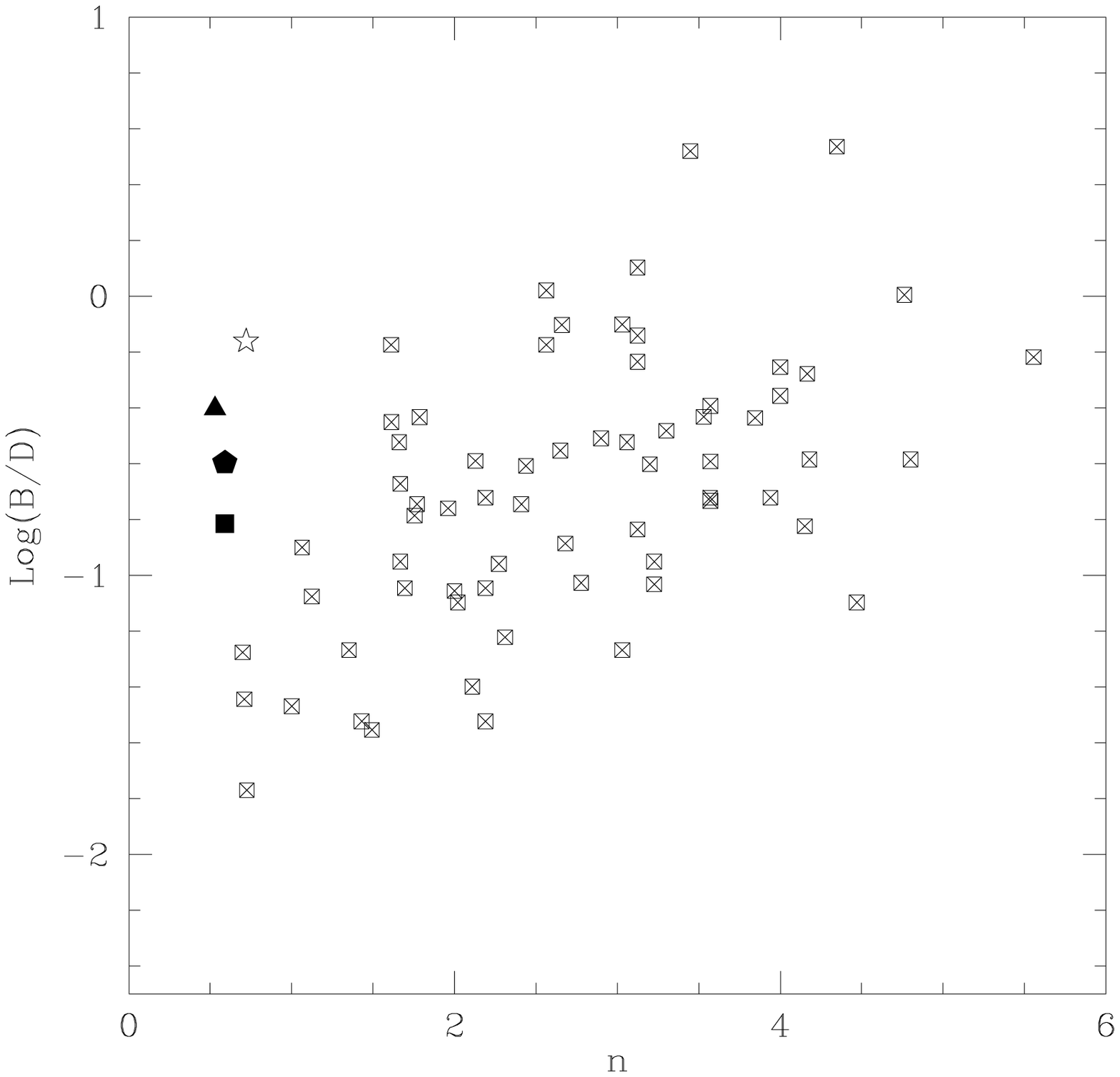}  
\caption{Left panel - B/D distribution for the PRGs of the sample 
(filled region), for early-type galaxies, by Bothun and Gregg 
(\cite{Bothun90}) (solid line) and for spiral galaxies, by
M\"{o}llenhoff \& Heidt (\cite{mollenhoff2001}); Khosroshahi et al. 
(\cite{khos2000b}).
Right panel - Bulge-to-Disk ratio as function of the n exponent, for
PRGs and for spiral galaxies. Code for symbols is the same adopted 
in Fig.\ref{bulge}.}  
\label{BD}
\end{figure*}

The {\it Bulge-to-Disk ratio (B/D)} for the host galaxy in PRGs of our sample 
falls in the range of values typical for disk-dominated S0 galaxies and for 
spiral galaxies, as shown in Fig.~\ref{BD} (left panel). 
The right panel of Fig.~\ref{BD} shows the correlation between the 
Bulge-to-Disk ratio (B/D) and the $n$ exponent: as stressed by 
M\"{o}llenhoff \& Heidt (\cite{mollenhoff2001}), for spiral galaxies the
B/D descreases with $n$. The host galaxy of PRGs are characterized by an higher
value of the B/D ratio with respect to the spiral galaxies, for the same value
of $n$. This is mostly due to the differences in the disks scalelenghts: 
disks in the PRG host galaxy are much smaller than disks in spiral galaxies 
(see Sec.~\ref{param_hg3}).

\section{Study of the light distribution in the polar  
structure}\label{param_pr}

An important quantity related to the size of the polar ring is the moment
of its radial distribution
\begin{equation}\label{eq_drr} 
(\Delta R)^2 = \frac{\int_{r_{min}}^{\infty}(r-\bar{R})^{2} \mu(r) dr}
{\int_{r_{min}}^{\infty} \mu(r) dr}
\end{equation}
where $\bar{R}$ is the average radius, weighted by the surface brightness 
distribution, given by
\begin{equation}
\bar{R} = \frac{\int_{r_{min}}^{\infty} r \mu(r) dr}{\int_{r_{min}}^{\infty} 
\mu(r)dr}
\end{equation}
and $r_{min}$ is equal to 3 times the effective radius of the central
component. 

For a pure exponential disk, the  $\Delta R / \bar{R}$ ratio tends to
unity when {\it r} goes to infinity. For a real object, this value is
expected to be less than 1, because of its finite extension.
This is confirmed by the  $\Delta R / \bar{R}$ values derived for a sample of
spiral galaxies (de Jong \cite{deJong96}) in the B band: this quantity varies
from $45\%$ to $75\%$ and the average value is $\Delta R /\bar{R} \sim 65\%$.   

\begin{figure}
\centering
\includegraphics[width=8cm]{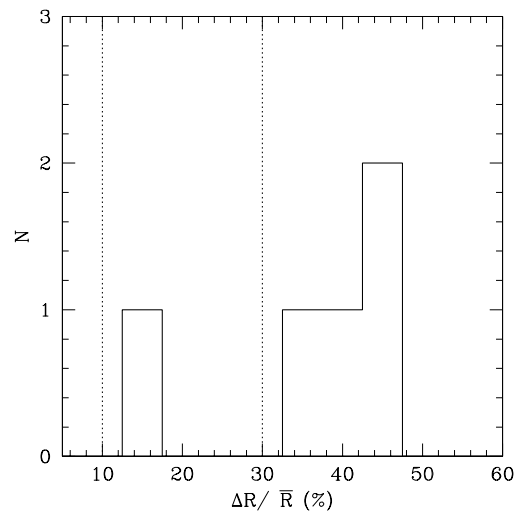}  
\caption{Distribution of the $\Delta R / \bar{R}$ ratio derived 
for each polar ring galaxy in the sample. The dotted lines limit the range of
$\Delta R / \bar{R}$ ratio for annuli in a quasi-equilibrium configuration,
formed in the accretion/merging scenario in the simulations reported by
Christodoulou et al. (\cite{chris92}) and Katz \& Rix \cite{Katz92}.}
\label{hist_drr}
\end{figure}

The $\Delta R / \bar{R}$ is a key parameter in the studies about polar ring 
stability (Sec.\ref{intro}).
In order to derive this quantity for every polar ring galaxy in the sample,  
a folded light profile was computed in the K band, for each object, 
from the surface brightness profiles extracted along the ring major axis 
(showed in Fig.4 and Fig.5, Paper I). 
For the polar ring galaxy ESO~415-G26 the average ring profile was
obtained in the B band, where this component is significantly brighter 
than in the K band (see Sec.7 in Paper I).
In Table~\ref{drr} we list the $\Delta R / \bar{R}$ ratio derived for each polar 
ring galaxy in the sample: this value is in principle a lower limit, 
since the polar ring light from regions
closer to the host galaxy is not included in the computation.
In Fig.~\ref{hist_drr} these values are compared with the typical 
$\Delta R / \bar{R}$ ratio for annuli in a quasi-equilibrium configuration, 
derived by Christodoulou et al. (\cite{chris92}) and 
Katz and Rix (\cite{Katz92})  from the stable configuration of PR 
hydrodynamical simulations.
All PRGs in this sample, but AM~2020-504, are outside
this range, and have larger $\Delta R / \bar{R}$ ratios.
Indeed, a larger number of simulations are needed to test these values for
stability, and verify whether those PRGs with a $\Delta R / \bar{R}$ close to
$30\%$ may be also considered as quasi-stable structures.

\begin{table}
\centering
\caption[]{$\Delta R / \bar{R}$ ratio for polar ring galaxies of the sample.} 
\label{drr} 
\begin{tabular}{cc}
\hline\hline
Object & $\Delta R / \bar{R}$ \% \\
\hline
A0136-0801 & 45\\
ESO 415-G26 & 47\\
ARP 230 & 35\\
AM 2020-504 & 16\\
ESO 603-G21 & 39 \\
\hline
\end{tabular}
\end{table}                                                                     

\section{Properties of the optical and NIR light distributions 
in PRGS}\label{sum}
Now we summarize which are the main observational properties of the PRGs  
studied in this work: we firstly discuss about the global characteristics
of the central host galaxy and then those relative to the polar structure.

\subsection{Is the central component a normal S0 galaxy?}
The morphology of the host galaxy in all PRGs of our sample, except for 
AM~2020-504, resembles that of an S0 galaxy, which is supported  by the
disk-like structure present along their major axis.
However, a detailed study of the light and color distributions
have shown that this component differs from a ``standard'' S0 galaxy.
The mean colors of the stellar
component outside the nucleus are, on average, bluer than the typical
values for elliptical and S0 galaxies; this is found in nearly all
objects.
Furthermore, in all systems a color gradient toward bluer colors is observed
from the nucleus to the outer regions. 
We found that most of the light comes from stars whose mean
age varies between $1$ and $3$ Gyrs, which is
significantly younger than the typical age for early-type galaxies.
The 2D model of the host galaxy light distribution, performed in the Kn band,
has shown that this component is reproduced by a nearly exponential
bulge and an outer exponential disk.
The analysis of their structural parameters suggests that
the host galaxy in PRGs is characterized by a more ``compact''  bulge
with respect to that in S0 galaxies, and by a brighter and smaller disk
than that of a typical S0's disk. In particular,
the PRGs bulge scale parameters, except for AM~2020-504, are similar to 
those of late-type galaxies rather than to those of early-type systems.
The resulting B/D ratio is in the range expected for disk-dominated S0 galaxies
and for spiral galaxies.
All PRG bulges, except for AM~2020-504, tends to be outside from 
the photometric plane which characterizes elliptical galaxies and the bulge of 
spiral galaxies. This evidence suggests that the host galaxy in 
all PRGs, except for AM~2020-504, may heve had a different formation history 
respect to Ellipticals and spiral bulges.

\subsection{Is the polar ring similar to a spiral disk?}

In all PRGs of our sample the colors of the polar structure fall in the same 
regions where late-type systems, dwarf and spiral galaxies, are also
found; they are, on average, bluer than the central host galaxy, 
implying a younger age for this component. 
In all PRGs studied in this work, the light of the polar
ring comes predominantly from stars not older than $1$ Gyr, which is
comparable with younger spiral galaxies.
An interesting result was derived by studying the light distribution of 
the polar structure in all objects of the sample: the ratio between its
radial extension ($\Delta R$) and its mean radius ($\bar R$) varies in the
range from  $35\%$ to $50\%$, with the exception of
AM~2020-504, whose polar ring is characterized by a significantly smaller
$\Delta R / \bar R$  (about $10\%$). The typical value of $\Delta R / \bar R$ 
for spiral galaxies varies from $45\%$ to $75\%$.
 
Previous studies on PRGs, including also the objects in our sample, have
shown that they are characterized by a very large amount of HI gas,
which is always associated to the polar structure (van Gorkom et
al. \cite{vgorkom87}; Arnaboldi et al. \cite{magda97}; van Driel et
al. \cite{vdriel2000}). The total amount of HI is especially high, when 
compared with the gas content of normal S0 galaxies, which often have no
detectable HI.
The HI mass in the polar ring galaxies studied in this work is larger than 
$10^9 M_\odot$ (see Sec.~7 in Paper I for more details). 
For a reasonable mass-to-light ratio ($M/L \sim 2$ in the
NIR, from Matthews et al. \cite{Matthews98}) we found that the total
baryonic mass (gas plus stars)  in the polar structure for all PRGs, but
AM~2020-504, is comparable with, or even 
higher than, the total luminous mass in the host galaxy (see
Table~\ref{tab_MB}).

\section{Conclusions}\label{conclu}
In this work we have presented the analysis of the detailed photometric study 
for a sample of PRGs, selected from the Polar Ring Catalogue (PRC,
Whitmore et al. \cite{PRC}), based on new NIR observations presented in
Iodice et al. \cite{paperI} (Paper~I).
We now wish to compare the properties predicted for PRGs in different  
formation scenarios, presented in  Sec.\ref{intro} (see also Iodice et
al. \cite{4650aI}), against the global properties observed for the polar
ring systems studied in this work.  

For all PRGs of our sample, except for AM~2020-504, 
the published simulations of the accretion/stripping scenario are 
so far not able to predict
\begin{enumerate}
\item the main characteristics in the light and color distribution of the  
host galaxy, which make this component different from a standard S0
system; 
\item the large values for the
$\Delta R / \bar R$ ratio, which is related to the ring extension;
\item the large baryonic mass (stellar + gas), shown in Table~\ref{tab_MB},
in the polar ring. 
\end{enumerate}

The observed properties for AM~2020-504 (see Sec.7 in Paper~I) suggest
that this polar ring may be the single case in our sample which may
be formed through an accretion or gas-stripping involving an elliptical
galaxy. 

All the observed properties of the host galaxy and polar structure can be
more easily explained by the dissipative merger scenario proposed by Bekki 
(\cite{Bekki98}).
In this scenario both the central S0-like system and ring component in a
polar ring galaxy are simultaneously formed through a dissipative merger
between two disk galaxies.  
The required constraints on the specific orbital configurations and gaseous 
dissipation in galaxy merging naturally explain the prevalence of S0-like 
systems among polar ring galaxies (e.g., Whitmore \cite{whitmore91}) and the 
appreciably  larger amount of interstellar gas in PRGs (van Gorkom et
al. \cite{vgorkom87}, 
Arnaboldi et al. \cite{magda97}, van Driel et al. \cite{vdriel2000}). 
This scenario does predict peculiar characteristics for the host galaxy:
the progenitor galaxy (the intruder) experiences both a heating of the disk
(it puffs up) and energy dissipation.
The energy dissipation leads to an higher increase of the mass density in the 
center, with respect to the unperturbed disk, which may develop a central  
small and nearly exponential bulge: this is very similar to what we have 
detected in nearly all PRGs of our sample. 
The evolutionary timescales of the merging process, which is about $10^9$
yr, is also consistent with the young age, predicted for PRGs in this
work, both for the host galaxy ($1$ to $3$ Gyr) and polar structure
($\sim 1$ Gyr). 
Furthermore,  the different morphologies observed for polar rings,
such as narrow rings (e.g. ESO~415-G26, or in ARP~230) and wide disk-like 
structures with no central hole (e.g. NGC~4650A, see Iodice et
al. \cite{4650aI} 
and Gallagher et al. \cite{4650aG}), are related to the orbital parameters of 
galaxy merging and the initial mass ratio of the two interacting galaxies.

An important constraint to the Bekki scenario is the small value of the
relative velocities ($V \sim 33$ km s$^{-1}$) that the two merging
galaxies need to have to form PRGs: such velocities are more likely to
occur in high redshift universe rather than nearby,  where bound group of
galaxies are virialized and therefore their relative 
velocities are larger. Reshetnikov (\cite{Resh97}) have found
an increasing rate of detection for PRGs toward higher redshift:
among all galaxy types, in the Hubble Deep Field (Williams et
al.  \cite{Williams95}) candidate polar ring galaxies are $\sim 0.7\%$,
while in the local universe this is $\sim 0.05\%$ (Whitmore et
al. \cite{PRC}).       
Although uncertainties in the numerical treatment of gas dynamics and star 
formation still remain in the Bekki's approach, dissipative galaxy merging, 
with specific initial conditions, seems now a promising scenario to 
to explain the formation of Polar Ring Galaxies and
their observational properties.

\begin{table}
\caption[]{Mass of the stellar component in the host galaxy (second column) and
the total baryonic mass in the polar structure (third column), which includes 
the mass of the stellar component and the mass of the gas in the form of  
neutral  (HI) and molecular hydrogen (HII). See Sec.7, Paper~I for
additional references.}
\label{tab_MB}
\begin{tabular}{lcc}
\hline\hline
Object & $M_{star}$ (HG) & $M_{gas} + M_{star}$ (PR)\\
       &  $10^9 M_{\odot}$ & $10^9 M_{\odot}$\\
\hline
ESO 415-G26 & 9 & 10\\
ARP 230 & 2 & 5\\
AM 2020-504& 6 & 5\\
ESO 603-G21& 2 & 10\\
\hline
\end{tabular}
\end{table}

\begin{acknowledgements}
The authors wish like to thank the referee, V. Reshetnikov, whose comments 
and suggestions helped to improve this work.
E.I and M.A. wish to thank Prof. Capaccioli and the staff of the
Observatory of Capodimonte for the help and support during the realization
of this work. E.I. and M.A. would like to thank G. De Lucia for the help
in the use of GISSEL, the stellar population synthesis model. 
\end{acknowledgements}


\begin{thebibliography}{}
\bibitem[1978]{Aar78} Aaronson, M., Cohen, J., Mould, J. and Malkan, M. 
1978, \aj, 223, 824  

\bibitem[1995]{magda95} Arnaboldi, M., Freeman, K. C., Sackett, P. D., Sparke, L. 
S. and Capaccioli, M., 1995, \pasp, 43, 1377

\bibitem[1997]{magda97} Arnaboldi, M., Oosterloo, T., Combes, F., Freeman, K. C. and 
Koribalski, B., 1997, \aj, 113, 585

\bibitem[1998]{Bekki98} Bekki, K. 1998, \apj, 499, 635
      
\bibitem[1997]{Bekki97} Bekki, K., 1997, \apj, 490, L37

\bibitem[1999]{Berg99} Bergvall, N., Ronnback, J., Masegosa, J. and Ostlin, G. 
1999, \aap, 341, 697                         

\bibitem[1990]{Bothun90} Bothun, G. D. and Gregg, M. D. 1990, \apj, 350, 73
                                                         
\bibitem[1984]{Bothun84} Bothun, G. D., Romanishin, W., Strom, S. E. and Strom, 
K. M. 1984, \aj, 89, 1300  
                                                                          
\bibitem[1982]{Bothun82} Bothun, G. D. 1982, \apjs, 50, 39 

\bibitem[1993]{BC93} Bruzual, G. and Charlot, S. 1993, \apj, 405, 538
                                                          
\bibitem[1993]{Caon93} Caon, N., Capaccioli, M. and Capaccioli, M. 1993, 
\mnras, 265, 1013                                 

\bibitem[1992]{Cap92} Capaccioli, M., Caon, N. and D'Onofrio, M., 1992, \mnras, 
259, 323                                                     

\bibitem[1996]{CWB96} Charlot, S., Worthey, G. and Bressan, A. 1996, \apj, 
457, 625                          

\bibitem[1992]{chris92} Christodoulou, D. M., Katz, N., Rix, H. W. and 
Habe, A. 1992, \apj, 395, 113                          

\bibitem[1988]{Davies88} Davies, J. I., Phillipps, S.,
Cawson, M. G. M., Disney, M. J. and Kibblewhite, E. J., 1988, \mnras, 232, 239  
           
\bibitem[1996]{deJong96} de Jong, R. S. 1996, \aaps, 118, 557   
            
\bibitem[1994]{deJong94} de Jong, R. S. and van der Kruit, P. C. 1994, \aaps, 
106, 451       

\bibitem[1995]{driver95} Driver, S.P., Windhorst R.A., Ostrander E.J.,
Keel W.C., Griffiths R.E. and Ratnatunga K.U., 1995, \apjl, 449, L23  

\bibitem[1985]{Frog85} Frogel, J. 1985, \apj, 298, 528 

\bibitem[1983]{Frog83} Frogel, J., Cohen, J. and Persson, E. 1983, \aj, 275, 773

\bibitem[2002]{4650aG} Gallagher, J.S., Sparke, L.S., Matthews, L.D.,
Frattare, L.M., English, J., Kinney, A.L., Iodice, E., Arnaboldi, M. 2002, 
\apj, 568, 199                      

\bibitem[1996]{Giov96} Giovanardi, C. and Hunt, L. K. 1996, \aj, 111, 1086
                       
\bibitem[1985]{Glass85} Glass, I. S. and Moorwood, A.F.M., 1985, \mnras, 214, 429
                                 
\bibitem[1997]{GCW97} Gordon, K. D., Calzetti, D. and Witt, A. N. 1997, \apj, 
487, 625                                       

\bibitem[1995]{Hibbard95} Hibbard, J. E. and Mihos, J. C., 1995, \aj, 110, 140 

\bibitem[2002a]{4650aI} Iodice, E., Arnaboldi M., De Lucia, G., Gallagher, J.S.,
Sparke, L.S. and Freeman, K.C. 2002, \aj, 123, 195

\bibitem[2002b]{paperI} Iodice, E., Arnaboldi M., 
Sparke, L.S., Gallagher, J.S. and Freeman, K.C. 2002, \aap, Paper I  
                   
\bibitem[1992]{Katz92} Katz, N. and Rix, H. 1992, \apjl, 389, L55

\bibitem[2000]{khos2000a}  Khosroshahi, H.G., Wadadekar, Y. and Kembhavi, A.,
and Mobasher, B., 2000, \apj, 531, L103  

\bibitem[2000]{khos2000b}  Khosroshahi, H.G., Wadadekar, Y. and Kembhavi, A.,
2000, \apj, 533, 162  

\bibitem[1983]{Kenn83}  Kennicutt, R. C., Jr., 1983, \apj, 272, 54  

\bibitem[1994]{Li94} Li, J.G. and Seaquist, E. R., 1994, \aj, 107, 1953

\bibitem[1998]{Matthews98} Matthwes, L.D., van Driel, W. and Gallagher, J.S. 1998, 
\aj, 116, 2196  

\bibitem[2001]{mollenhoff2001} M\"{o}llenhoff, C. and Heidt, J. 2000, 
\aap, 368, 16  

\bibitem[1979]{Persson79} Persson, S. E., Frogel, J. A. and Aaronson, M. 1979, 
\apjs, 39, 61  

\bibitem[1997]{Resh_Sot97} Reshetnikov, V. and Sotnikova, N. 1997, \aap, 325, 933
                                   
\bibitem[1997]{Resh97}  Reshetnikov, V.P. 1997, \aap, 321, 749 

\bibitem[1996]{Resh96} Reshetnikov, V. P.,
Hagen-Thorn, V. A. and Yakovleva, V. A., 1996, \aap, 314, 729

\bibitem[1994]{Resh94} Reshetnikov, V. P.,
Hagen-Thorn, V. A. and Yakovleva, V. A., 1994, \aap, 290, 693

\bibitem[1991]{Sackett91} Sackett. P., 1991, in 
Warped Disks and Inclined Rings around Galaxies, ed. S. Casertano, P. Sackett 
and F. Briggs (New York: Cambridge Univ. Press), p.73         

\bibitem[1955]{Salp55} Salpeter, E. E. 1955, \apj, 121, 161 

\bibitem[1983]{Schweizer83} Schweizer, F., Whitmore, B. C. and Rubin, V. C., 1983,
\aj, 88, 909

\bibitem[1980]{Shane80} Shane, W.W., 1980, \aap, 82, 314   

\bibitem[1991]{Sparke91} Sparke, L.S., 1991 in
Warped Disks and Inclined Rings around Galaxies,
ed. S. Casertano, P. Sackett and F. Briggs (New York: Cambridge Univ. Press), 
p.85  

\bibitem[1985]{Thuan85} Thuan, T.\ X.\ 1985, \apj, 299, 881          

\bibitem[1977]{Toomre77} Toomre, A., 1977, \araa, 15, 437  
                                                             
\bibitem[2000]{vdriel2000} van Driel, W., Arnaboldi, M.,
 Combes, F. and Sparke, L. S., 2000, \aaps, 141, 385

\bibitem[1987]{vgorkom87} van Gorkom, J. H., Schechter, P.L.
and Kristian, J., 1987, \apj, 314, 457

\bibitem[1991]{whitmore91} Whitmore, B. C. 1991, in Warped Disks and Inclined Rings
around Galaxies, ed. S. Casertano, P. Sackett and F. Briggs 
(New York: Cambridge Univ. Press), 60

\bibitem[1990]{PRC} Whitmore, B. C., Lucas, R. A., McElroy, D. B., 
Steiman-Cameron, T. Y.,Sackett, P. D. and Olling, R. P. 1990, \aj, 100, 1489

\bibitem[1995]{Williams95} Williams, R., Dickinson, M.,
Giavalisco, M., Gilliland, R., Ferguson, H., Fruchter, A., McElroy, D., Lucas, R.,
Petro, L., Postman, M., American Astronomical Society Meeting, 187 

\end{thebibliography}
\end{document}